\documentclass{elsart1p}
\usepackage{graphicx}
\usepackage{amssymb}
\usepackage{amsmath}
\begin{document}
\begin{frontmatter}
%
%
%
%
%
\title{Extended Nambu--Jona-Lasinio Model with covariant regularization}
\author[Coimbra]{J. Moreira}
\author[Coimbra]{B. Hiller}
\author[Coimbra,Dubna]{A. A. Osipov}
\author[Coimbra]{A. H. Blin}
\address[Coimbra]{Centro de F\'{i}sica Computacional, Departamento de F\'{i}sica da Universidade de Coimbra, 3004-516 Coimbra, Portugal}
\address[Dubna]{On leave from Dzhelepov Laboratory of Nuclear Problems, Joint Institute for Nuclear Research, 141980 Dubna, Moscow Region, Russia}

\begin{abstract}
Several advantages in the use of a Pauli-Villars regularization  procedure in extended Nambu--Jona-Lasinio models with Polyakov loop are discussed.
\end{abstract}

\end{frontmatter}

\section{Introduction}
\label{}


It has been shown that the extension of the Nambu--Jona-Lasino Model in the light quark sector (u, d and s) to include the 't Hooft determinantal interaction (NJLH) which explicitly breaks the unwanted axial symmetry introduces a ground state stability problem \cite{Osipov:2006} which can be solved by the addition of general non-derivative spin-zero eight quark interaction terms (NJLH8) \cite{Osipov:2005b}. The model parameters can be fitted in such a way that the low lying scalar and pseudoscalar spectra are left relatively unchanged (apart from a decrease in the sigma meson mass) throughout a wide range of values for the OZI-violating part of these 8q interactions \cite{Osipov:2006a}. They were however shown to have a significant impact on the position of the critical endpoint where the transition goes from crossover to first order (the CEP is moved to lower chemical potential and increasing temperature with stronger 8q interactions) \cite{Hiller:2010} as well as the temperature at which the transition occurs (lowered for stronger 8q interactions) \cite{Osipov:2007b}. The extension of this model to include the Polyakov loop (PNJL) can be done straightforwardly \cite{Moreira:2010bx} and enables the simultaneous study of chiral restoration and deconfinement (at least approximately).

The transition between the confined-deconfined phases is driven by the temperature dependence of the additional pure gluonic term, the  Polyakov potential, $\mathcal{U}$, for which several forms have been proposed. In \cite{Weise:2006} a polynomial form motivated by a Ginzburg-Landau \emph{ansatz} was used.
 In \cite{Roessner:2006xn} a logarithmic term inspired by the Haar measure of $SU(N_c)$ group integration is introduced. In \cite{Fukushima:2008pe} this logarithimic term is combined with an exponential term derived in the strong coupling expansion of the lattice QCD action, and in \cite{Bhattacharyya:2010wp} the polynomial and logarithmic forms are combined (for details see these references and our discussion thereof in \cite{Moreira:2010bx} where we also discuss the parametrization).
 
Integrating the gap equations selfconsistently with the stationary phase equations,
\begin{align}
\label{staeq}
\left\{
\begin{array}{l}
m_u-M_u=G h_u +\frac{\kappa}{16}h_d h_s +\frac{g_1}{4}h_u h^2_f+\frac{g_2}{2}h_u^3\\
m_d-M_d=G h_d +\frac{\kappa}{16}h_u h_s +\frac{g_1}{4}h_d h^2_f+\frac{g_2}{2}h_d^3\\
m_s-M_s=G h_s +\frac{\kappa}{16}h_u h_d +\frac{g_1}{4}h_s h^2_f+\frac{g_2}{2}h_s^3
\end{array}
\right. ,
\end{align}
(where $m_f$ and $M_f$ are the current and dynamical masses respectively), the thermodynamical potential is obtained
\cite{Hiller:2010} ($T$ is the temperature, $\mu$  the chemical potential, $G$, $\kappa$, $g_1$ and $g_2$ are the coupling strengths of the of the NJL, 't Hooft, OZI-violating and non-violating 8-quark interactions):
\begin{align}
\label{Omega}
{\Omega\left(M_f,T,\mu,\phi,\overline{\phi}\right)}
=& \frac{1}{16}\left.\left(4Gh_f^2+\kappa h_uh_dh_s+\frac{3g_1}{2}\left(h_f^2\right)^2+3g_2h_f^4\right)\right|_0^{M_f} \nonumber \\
+&\frac{N_c}{8\pi^2}\!\sum_{f=u,d,s}\!\!\left( {J_{-1}(M_f^2,T,\mu,\phi,\overline{\phi} )} + C(T,\mu )\right)+ {\mathcal{U}\left(\phi,\overline{\phi},T\right)}.
\end{align}
The effect of the Polyakov loop was included straightforwardly by noting that its phase enters the action as an imaginary $\mu$ thus resulting in the following generalizations:
\begin{align}
\label{nfdefs}
\tilde{n}_{q(\overline{q})}\left(E_p,\mu,T,\phi,\overline{\phi}\right)&\equiv
\frac{1}{N_c}\sum_{i}^{N_c}n_{q(\overline{q})} (E_p,\mu+\imath \left(A_4\right)_{ii},T),
\nonumber\\
f^{+(-)}\left(E_p,\mu,T,\phi,\overline{\phi}\right)&\equiv Log \prod^{3}_{i=1}
\frac{e^{-\left(E_p\mp\mu\right)/T}}{n_{q(\overline{q})} (E_p,\mu+\imath \left(A_4\right)_{ii},T)}.
\end{align}
Here $E_p=|\overrightarrow{p}_E|^2+M^2$ and the (anti-)quark occupation numbers are give as usual by: 
$n_{q(\overline{q})}=(1+e^{(E_p\mp\mu)/T})^{-1}$

The use of a covariant regulator with two Pauli-Villars (PV) subtractions in the momentum integrand function ($\hat{\rho}_{\Lambda\vec{p}_E}=
 1-\left(1-\Lambda^2 \partial_{\vec{p}_E^{\, 2}}\right)\exp\left(\Lambda^2\partial_{\vec{p}_E^{\, 2}}\right)$) results in the correct asymptotic behavior of several thermodynamic quantities when $T\rightarrow\infty$ \cite{Osipov:2007b,Moreira:2010bx}. This feature that had been reproduced using a 3D momentum cutoff ($\hat{\rho}^{3D}=\Theta\left(\Lambda-\left|\overrightarrow{p_E}\right|\right)$) only by eliminating the cutoff in the convergent parts of the relevant integrals,  can be done here while consistently maintaining the cutoff over all contributions. Here we show that using this procedure several undesirable features are eliminated, such as the deviation of the asymptotic solution for the Polyakov loop from the value dictated by the pure gluonic term that is added to the potential (the Polyakov potential), or the feature that the dynamical mass of the quark is going below the current mass value, or the inability to obtain the Stefan-Boltzmann asymptotic limit for pressure as a function of the temperature. 

The vacuum and medium contributions can be separated and depending on the choice of the regularization procedure we obtain:
\begin{align}
J^{vac (3D)}_{-1}(M^2)
=&\Lambda\left(2\Lambda^3-\sqrt{M^2+\Lambda^2}\left(M^2+2\Lambda^2\right)\right)
+M^4 \mathrm{ArcSinh}\frac{\Lambda}{M}\nonumber\\
J^{vac (PV)}_{-1}(M^2)
=&\frac{M^4-\Lambda^4}{2}\ln(1+\frac{M^2}{\Lambda^2})
-\frac{M^2}{2}\left(\Lambda^2+M^2\ln\frac{M^2}{\Lambda^2}\right)\\
J^{med (3D)}_{-1}(M^2,T,\mu,\phi,\overline{\phi})
=&-\int^\Lambda_0\mathrm{d}|\overrightarrow{p_E}|~
8|\overrightarrow{p_E}|^2 T
(f^+_{M}+f^-_M-(f^+_0+f^-_0))\nonumber\\
J^{med(PV)}_{-1}(M^2,T,\mu,\phi,\overline{\phi}),
=&-\int^{\infty}_0\mathrm{d}|\overrightarrow{p}_E| \frac{8|\overrightarrow{p}_E|^4}{3}\hat{\rho}_{\Lambda\vec{p}_E}
\left(
\frac{n_{q~M}+n_{\overline{q}~M}}{E_p}-
\frac{n_{q~0}+n_{\overline{q}~0}}{|\overrightarrow{p}_E|}
\right).
\nonumber
\end{align}
We use the notation with the subscripts $M$ and $0$ denoting the quantities defined in (\ref{nfdefs}) evaluated at those values for the mass (for the $M=0$ case we also set $\phi=\overline{\phi}=1$)  \cite{Hiller:2010}\cite{Moreira:2010bx}.

The constants of integration, $C(T,\mu)$, resulting from the integration of the gap equations over the mass are chosen as to counterbalance the part stemming from the zero-mass limit of integration and are therefore given by (note that while this may not be the standard way used to derive the 3D result the final result is equivalent to the usual one):
\begin{align}
 C^ {3D}(T,\mu)=&-\int^\Lambda_0\mathrm{d}|\overrightarrow{p_E}|8 |\overrightarrow{p_E}|^2 T 
 \left(f^+_0+f^-_0\right),\nonumber\\
C^{PV}(T,\mu) = &-\int^{\infty}_0\mathrm{d}|\overrightarrow{p}_E \frac{8|\overrightarrow{p}_E|^4}{3}
\left(\frac{n_{q~0}+n_{\overline{q}~0}}{|\overrightarrow{p}_E|}\right).
\end{align}
 
\section{Results}
Here we only present some selected results which illustrate the key points, for a more complete exposition and further details we refer to \cite{Moreira:2010bx}.
We tested the above mentioned four different regularization procedures (PV/3D, with/without regularization of the convergent medium contribution) with the four mentioned Polyakov potential forms, using parameter sets with and without eight quark interactions. The analysis of the normalized pressure difference at vanishing chemical potential which serves as a measure for the effective degrees of freedom,
$\nu(T)=(p(T)-p(0))/(\pi^2T^4/90)$, reveals that the failure to reach the Stefan-Boltzmann limit only happens in the case where we use the 3D cutoff everywhere (see Fig. 1a).

The removal of the cutoff in the medium part results in the dynamical mass dropping below the current mass going asymptotically 
to zero with the temperature increase (which happens for all the mentioned potential forms, see Fig. 1b for an example) as well as an overshooting of the asymptotic solution dictated by the pure gluonic part in the case of the potentials taken from \cite{Weise:2006} (see Fig. 1c), \cite{Fukushima:2008pe} and \cite{Bhattacharyya:2010wp} (in \cite{Roessner:2006xn} the logarithmic divergence prevents this from happening). This overshooting does not occur if the cutoff is kept (see Fig. 1c and 1d for the $\mu=0$ and $\mu>0$ cases respectively).

We can trace back the reason for these behaviours to the asymptotic behaviour of the derivatives of $J_{-1}$ with respect to $M$, $\phi$ and $\overline{\phi}$ when $T\rightarrow\infty$: the derivative with respect to the mass diverges as $T^2$ upon removal of the cutoff whereas it goes to zero when the cutoff is removed; the derivative with respect to $\phi$ ($\overline{\phi}$) diverges with $T^4$ upon the removal of the cutoff, the same order as the derivative of $\mathcal{U}$, originating a deviation from the solution dictated by the latter. The divergence is lower with cutoff and in this case the asymptotic solution is dictated by the Polyakov potential.

\begin{figure}[htp]
\centering
\begin{minipage}{\columnwidth}
$\begin{array}{cccc}
\includegraphics[width= 0.24 \columnwidth]{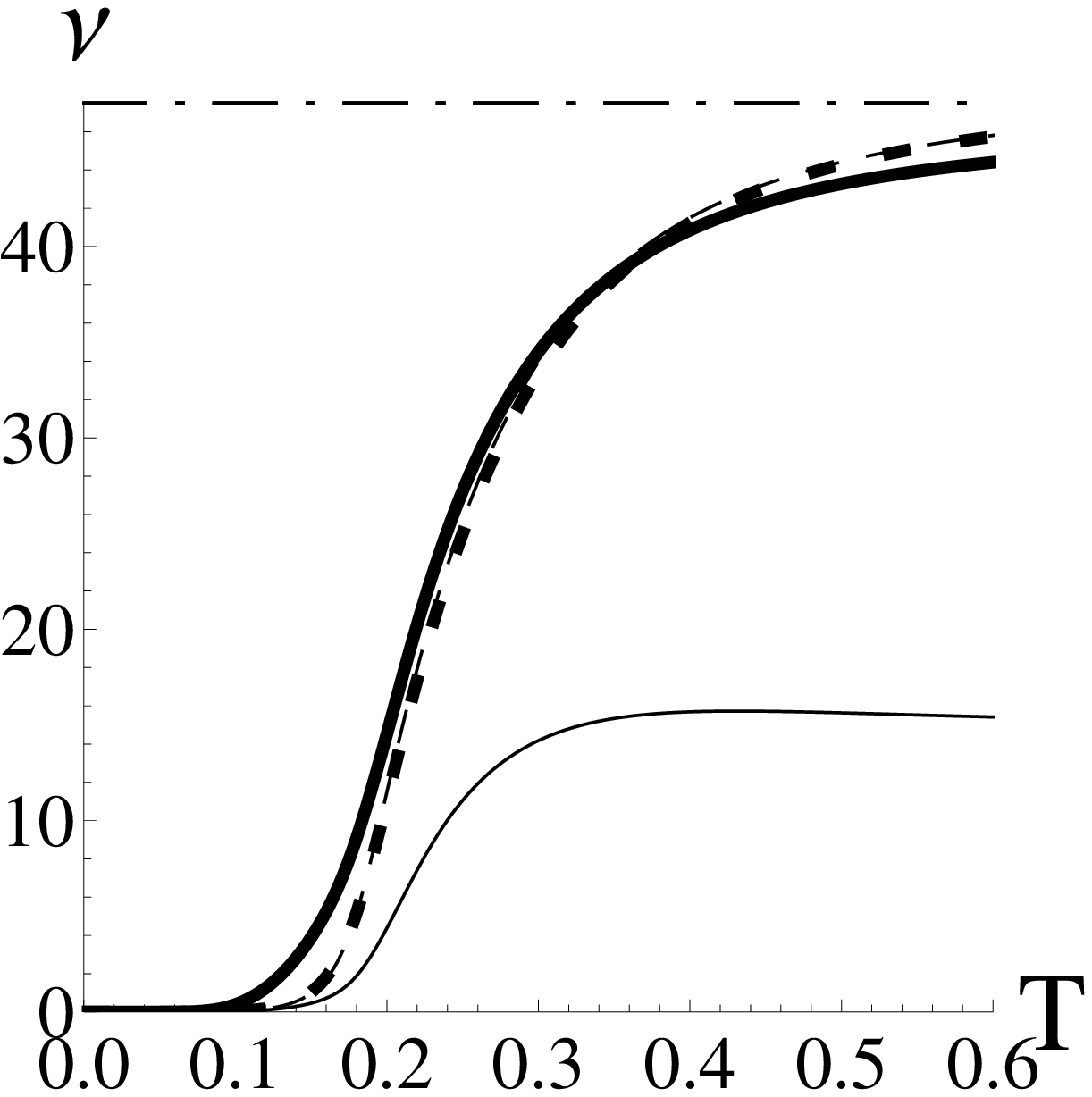}&
\includegraphics[width= 0.24 \columnwidth]{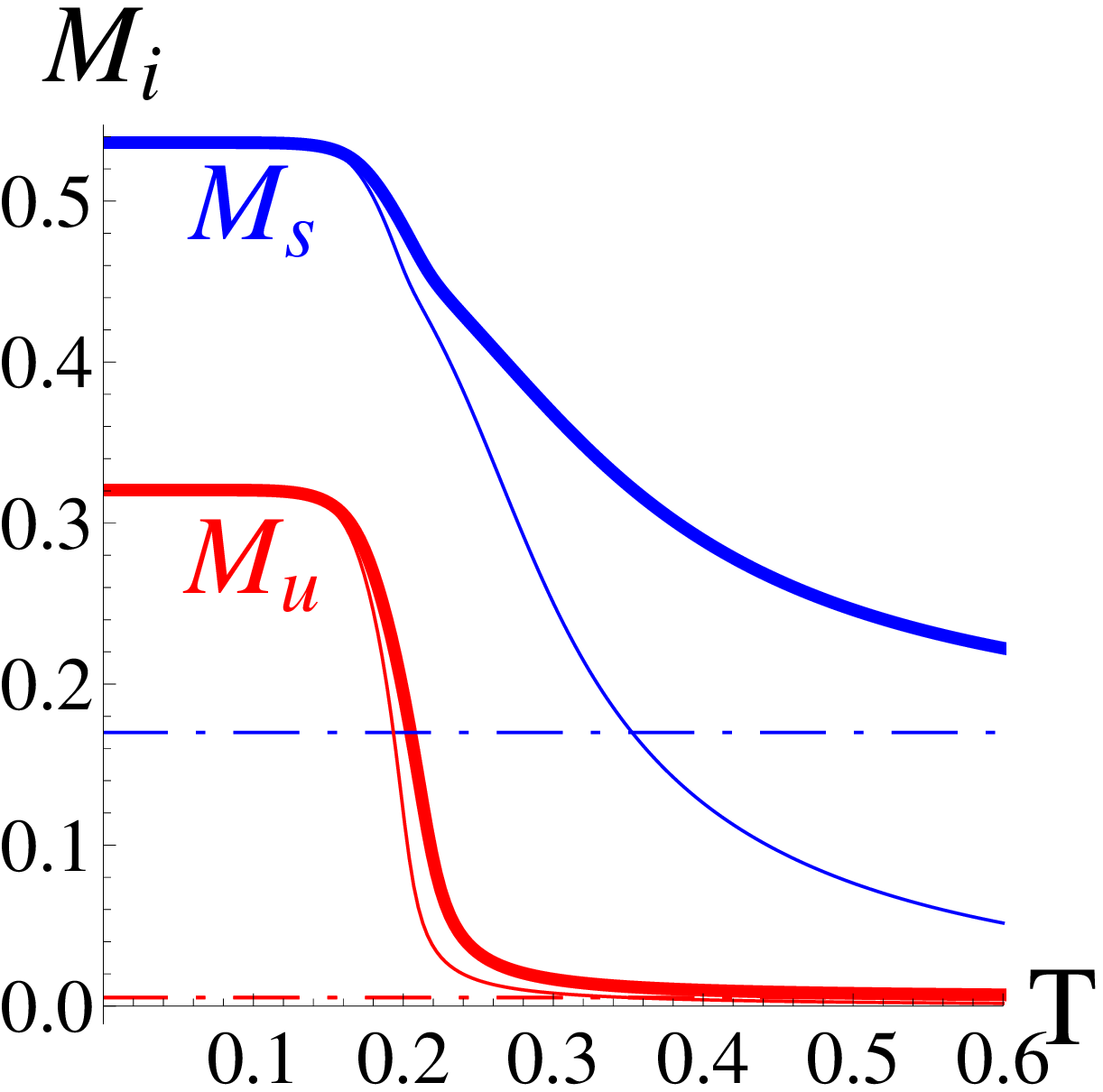}&
\includegraphics[width= 0.24 \columnwidth]{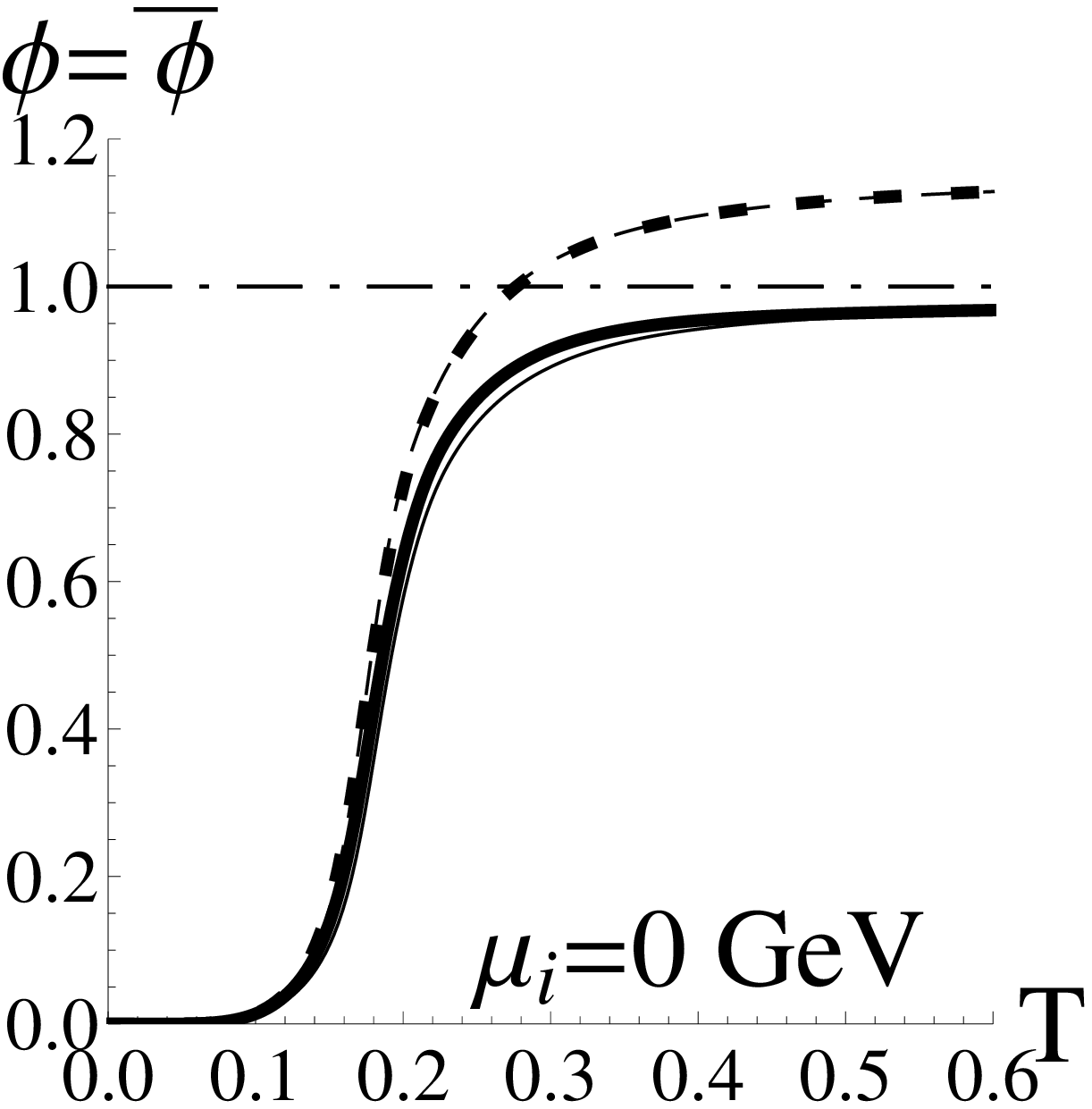}&
\includegraphics[width= 0.24 \columnwidth]{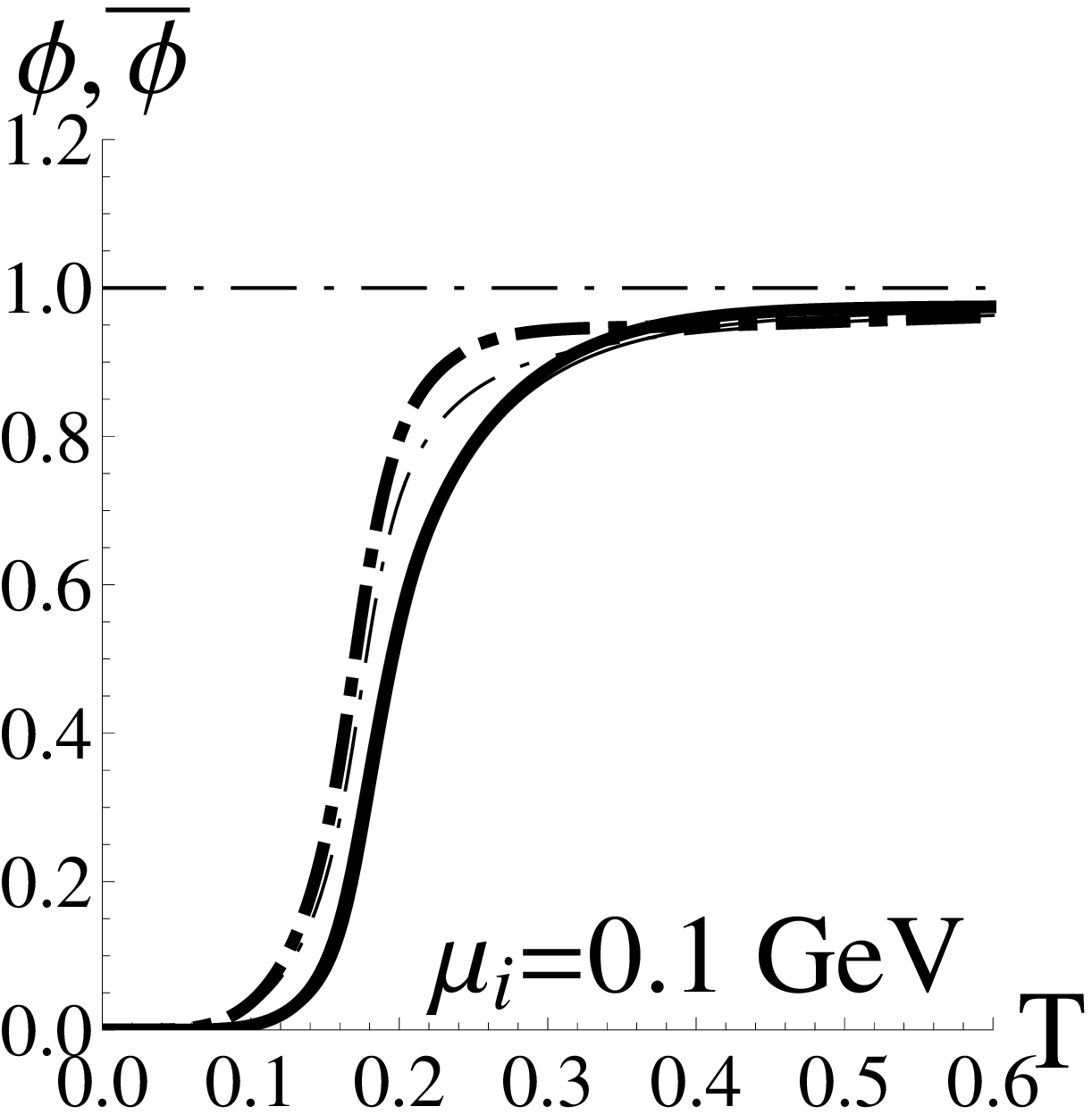}\\
 a) & b) & c) & d)
\end{array}$
\end{minipage}
\caption{Temperature dependence ($\left[T\right]=\mathrm{GeV}$) in the PNJLH model with $\mathcal{U}$ from \cite{Weise:2006} of: a) the number of effective degrees of freedom, 
$\nu(T)$ (thick lines using PV and thin using 3D, dashing denotes the removal of the cutoff in the medium contributions); dynamical mass  of the quarks  ($\left[M_i\right]=\mathrm{GeV}$) using PV regularization (thinner lines correspond to the removal of the cutoff in the medium; 3D results are qualitatively similar);
c) $\phi$ at $\mu=0$ ($\phi=\overline{\phi}$) the thick lines refer to the use of PV and thin to 3D (dashed with only the vacuum regularized); d) $\phi$ in full lines and $\overline{\phi}$ in dot-dashed lines at $\mu=100~\mathrm{MeV}$, thick and thin lines refer to PV and 3D respectively (only the cases with cutoff everywhere are displayed).}
\end{figure}

\section{Conclusions}
These qualitative features appear to be independent of the choice of parametrization (both of the quark interactions and the Polyakov potential) and are in fact a result of the regularization. For the studied quantities the choice of PV regularization with the cutoff kept over all contributions achieves the best results.


This  work  has  been  supported  in  part  by  grants  of Funda\c{c}\~{a}o para a Ci\^{e}ncia e Tecnologia,  FEDER,  OE,
SFRH/BPD/63070/2009, and Centro de F\'{i}sica Computacional, unit 405. We acknowledge  the support of the  European Community-Research  Infrastructure  Integrating  Activity  Study  of Strongly  Interacting  Matter  (acronym  HadronPhysics2, Grant   Agreement   No.   227431)   under   the   Seventh Framework Programme of the EU.


\begin{thebibliography}{00}
\bibitem{Osipov:2006} A.A. Osipov, B. Hiller, V. Bernard, A.H. Blin, Ann. of Phys. {\bf 321}, 2504 (2006); hep-ph/0507226.
\bibitem{Osipov:2005b} A.A. Osipov, B. Hiller and J.da Provid\^encia, Phys. Lett. B {\bf 634}, 48 (2006); hep-ph/0508058.
\bibitem{Osipov:2006a} A.A. Osipov, B. Hiller, A.H. Blin and J. da Provid\^encia, Ann. of Phys. {\bf 322}, 2021 (2007); hep-ph/0607066.
\bibitem{Hiller:2010} B. Hiller, J. Moreira, A.A. Osipov, A.H. Blin, Phys. Rev. D {\bf 81}, 116005 (2010);
        0812.1532 [hep-ph].
\bibitem{Osipov:2007b} A.A. Osipov, B. Hiller, J. Moreira, A.H. Blin, 
        J.da Provid\^encia, Phys. Lett. B {\bf 646}, 91 (2007); hep-ph/0612082.
\bibitem{Moreira:2010bx}
  J.~Moreira, B.~Hiller, A.~A.~Osipov and A.~H.~Blin,
  arXiv:1008.0569 [hep-ph].
\bibitem{Weise:2006} C. Ratti, M.A. Thaler, and W. Weise, Phys. Rev. D
        {\bf 73}, 014019 (2006).
\bibitem{Roessner:2006xn} S. Roessner, C. Ratti, W. Weise, Phys. Rev. D {\bf 75}, 034007 (2007).
\bibitem{Fukushima:2008pe} K. Fukushima, J. Phys. G {\bf 35}, 104020 
        (2008); arXiv:0806.0292 [hep-ph].
\bibitem{Bhattacharyya:2010wp}A. Bhattacharyya, P. Deb, S. K. Ghosh, R. Ray, Phys. Rev. D {\bf 82}, 014021 (2010),
        arXiv:1003.3337v1 [hep-ph]. 
         A. Bhattacharyya, P. Deb, A. Lahiri, R. Ray, arXiv:1010.2394 [hep-ph]
\end{thebibliography}
\end{document}